\documentclass[aps,showpacs,showkeys]{revtex4}

\begin{document}


\newcommand\beq{\begin{equation}}
\newcommand\eeq{\end{equation}}
\newcommand\beqa{\begin{eqnarray}}
\newcommand\eeqa{\end{eqnarray}}
\newcommand\ket[1]{|#1\rangle}
\newcommand\bra[1]{\langle#1|}
\newcommand\scalar[2]{\langle#1|#2\rangle}
\newcommand\ua{\uparrow}
\newcommand\da{\downarrow}

\newcommand\jo[3]{\textbf{#1}, #2 (#3)}


\title{\Large\textbf{Criterion for faithful teleportation with an
 arbitrary multiparticle channel}}

\author{Chi-Yee Cheung} \email{cheung@phys.sinica.edu.tw}
\affiliation{Institute of Physics, Academia Sinica, Taipei,
Taiwan 11529, Republic of China}

\author{Zhan-Jun Zhang} \email{zjzhang@ahu.edu.cn}
\affiliation{School of Physics \& Material Science, Anhui
University, Hefei 230039, China}


\begin{abstract}
We consider quantum teleportation when the given
entanglement channel is an arbitrary multiparticle state. A
general criterion is presented, which allows one to judge
if the channel can be used to teleport faithfully an
arbitrary quantum state of a given dimension. The general
protocol proposed here is much easier to implement
experimentally than the others found in the literature.
\end{abstract}

\pacs{03.67.Ac, 03.67.Hk, 03.67.Mn}

\keywords{quantum teleportation, multiparticle
entanglement, teleportation protocol}

\maketitle


Quantum teleportation is arguably the most novel
application of quantum mechanics in quantum information
science. This protocol provides a means of recreating an
arbitrary quantum state at a remote site without the need
of transferring any particles or a large amount of
classical information. The magic of teleportation is made
possible by prior quantum entanglement between the sender
(Alice) and the receiver (Bob). It is well known that, in
the original protocol proposed by Bennett et al.
\cite{Bennett-93}, if Alice and Bob share a two-qubit
entangled state (Bell or EPR state), then Alice can
teleport an arbitrary one-qubit state to Bob. This protocol
is linear, that means if Alice and Bob shares $N$ Bell
states, then Alice will be able to teleport an arbitrary
$N$-qubit state to Bob. In recent years, quantum
teleportation has been experimentally realized in several
different quantum systems
\cite{Marcikic-03,Riebe-04,Barrett-04,Zhao-04,Ursin-04}.

To facilitate the ensuing discussions, we first demonstrate
the teleportation of an arbitrary $N$-qubit state below.
The four Bell states are given by
 \beqa
 &&\ket{\phi^{1}}_{ab}={1\over\sqrt{2}}
 \Big(\ket{01}_{ab}-\ket{10}_{ab}\Big),\\
 &&\ket{\phi^{2}}_{ab}={1\over\sqrt{2}}
 \Big(\ket{01}_{ab}+\ket{10}_{ab}\Big),\\
 &&\ket{\phi^{3}}_{ab}={1\over\sqrt{2}}
 \Big(\ket{00}_{ab}-\ket{11}_{ab}\Big),\\
 &&\ket{\phi^{4}}_{ab}={1\over\sqrt{2}}
 \Big(\ket{00}_{ab}+\ket{11}_{ab}\Big).
 \eeqa
Suppose Alice shares a Bell state $\ket{\phi^1}_{ab}$ with
Bob (Alice holds qubit $a$ and Bob holds qubit $b$), and in
addition she owns an arbitrary $N$-qubit pure state
$\ket{\Psi}_{12...N}$ to be teleported to Bob. Note that if
the $N$ qubits are in a mixed state, then it can always be
purified by introducing ancilla qubits \cite{Chuang-02}.
However since the ancillas do not participate in the
teleportation process, we will ignored their possible
existence in the following discussions.

The product of $\ket{\Psi}_{12...N}$ and
$\ket{\phi^1}_{ab}$ can be rewritten as
 \beqa
 \ket{\Psi}_{12...N}\ket{\phi^1}_{ab}
 =-{1\over 2}\Big(\sum_{i=1}^4\ket{\phi^i}_{1a}U^i_{b}\Big)
 \ket{\Psi}_{b2...N},
 \label{teleport-N}
 \eeqa
where $\{U^i\}=\{I,\sigma^z,-\sigma^x,i\sigma^y\}$, and $I$
is the $2\times 2$ unit matrix. Therefore if Alice makes a
Bell state measurement on the qubit pair $(1,a)$ and sends
a 2-bit classical message to inform Bob of the outcome
($i$), then Bob can reconstruct the original $N$-qubit
state by applying a local unitary operation $U_b^i$ on his
qubit. The only difference is that qubit 1 has been renamed
$b$ and is now in his possession. It is easy to see that,
with more Bell states, this process can be repeated on the
other qubits in $\ket{\Psi}_{12...N}$. Therefore if Alice
shares $N$ pairs of Bell states with Bob, then she will be
able to teleport perfectly the entire state
$\ket{\Psi}_{12...N}$ to Bob.

However in practice the channel state shared by Alice and
Bob may not be a tensor product of $N$ Bell states. Then
one must consider each case individually to decide if it is
useful for teleportation, and if so how to proceed; there
exists no general rule. Some special cases have been
studied in the literature
\cite{Yeo-06,Chen-06,Lee-02,Rigolin-05,Agrawal-06,Zha-07,Mura-08},
and most of them are concerned with four-qubit channels.
For example, Yeo and Chua \cite{Yeo-06} introduced a
so-called ``genuinely four-qubit entangled state" which is
not reducible to a pair of Bell states, and showed that it
could be used to teleport an arbitrary two-qubit state.
Chen et al. \cite{Chen-06} generalized the results of
\cite{Yeo-06} to $N$-qubit teleportation. Rigolin
\cite{Rigolin-05} constructed 16 four-qubit entangled state
which are useful for two-qubit teleportation. Agrawal and
Pati considered teleportation using asymmetric W states.
Muralidharan and Panigrahi \cite{Mura-08} employed
a``genuinely entangled" channel of five qubits to teleport
two qubits. A criterion has been proposed by Zha and Song
\cite{Zha-07} in terms of the unitarity property of a
``transformation matrix", which tells if a four-qubit
entanglement channel supports two-qubit teleportation.
However no general results exist in the literature when the
quantum channel in question is an arbitrary multiparticle
entangled state.

In this paper, we consider the most general situation in
which Alice and Bob share an arbitrary bipartite state
$\ket{X}_{A_1...A_mB_1...B_n}$ of $(m+n)$ qubits, of which
$m$ qubits belong to Alice and $n$ to Bob. We shall first
assume $m\ge n$; the $m<n$ case can later be included in a
straightforward manner. As mentioned before, entanglement
is the key ingredient which makes quantum teleportation
possible. For an arbitrary bipartite state
$\ket{X}_{A_1...A_mB_1...B_n}$ (or $\ket{X}_{AB}$ for
short), the degree of entanglement between Alice's and
Bob's subsystems can be quantified by the von Neumann
entropy of either of two subsystems
\cite{Bennett-96,Popescu-97}, which is given by
 \beq
 E_{AB}=-\textrm{Tr}\,(\rho_A\textrm{log}_2\rho_A)
 =-\textrm{Tr}\,(\rho_B\textrm{log}_2\rho_B),
 \eeq
where $\rho_A$ and $\rho_B$ are the reduced density
matrices of the subsystems,
 \beqa
 \rho_A &=& \textrm{Tr}_B(\ket{X}_{AB}\bra{X}_{AB}),\\
 \rho_B &=& \textrm{Tr}_A(\ket{X}_{AB}\bra{X}_{AB}).
 \eeqa
Since we assume $m \ge n$, therefore
 \beq
 E_{AB}\le n.
 \eeq
Consider first the case
 \beq
 E_{AB}=n \label{maximal}
 \eeq
so that the entanglement between the two subsystems is
maximal. Then we must have
 \beq
 \rho_B=I_B/2^n,
 \eeq
where $I_B$ is the $2^n\times 2^n$ identity matrix in Bob's
Hilbert space $H_B$. Now consider another situation in
which Alice and Bob share $n$ pairs of Bell states, and in
addition Alice owns an arbitrary pure state $\ket{\cal{O}}$
of $(m-n)$ qubits. The combined $(m+n)$-qubit state is
 \beq
 \ket{\Lambda}_{AB}=\Big(\prod_{i=1}^n
 \ket{\phi^k}_{A_iB_i}\Big)\ket{{\cal O}}_{A_{n+1}...A_m},
 \label{Lambda}
 \eeq
where $k\in \{1,2,3,4\}$. Although the properties of
$\ket{{\cal O}}_{A_{n+1}...A_m}$ are irrelevant, for
definiteness and without loss of generality, we may take
 \beq
 \ket{{\cal O}}_{A_{n+1}...A_m}
 =\prod^m_{j=n+1}\ket{0}_{A_j}.
 \eeq
The reduced density matrix on Bob's side is given by
 \beq
 \tilde\rho_B=\textrm{Tr}_A\,\ket{\Lambda}_{AB}
 \bra{\Lambda}_{AB}=I_B/2^n.
 \eeq
It follows that $\ket{\Lambda}_{AB}$ and $\ket{X}_{AB}$ are
two purifications of the state $I_B/2^n$ in $H_B$ to the
joint space $H_B\otimes H_A$. By the freedom of
purification \cite{Hughston-93}, these two pure states are
related by an unitary transformation on Alice's side. In
other words, Alice can transform $\ket{X}_{AB}$ into
$\ket{\Lambda}_{AB}$ by applying a local unitary operation
${\cal U}_A$ on her qubits,
 \beq
 \ket{\Lambda}_{AB}={\cal U}_A\,\ket{X}_{AB},\label{UA}
 \eeq
where ${\cal U}_A$ is a $m$-qubit unitary operator in
$H_A$. It is important to note that Alice can carry out
this transformation by herself, and there is no need for
Bob to do anything. Note that, if $m<n$, then maximal
entanglement means $E_{AB}=m$ and Bob must carry out the
corresponding $n$-qubit transformation ${\cal U}_B$. And if
$m=n$, either party can do it.

Eq. (\ref{UA}) essentially establishes that, for any
arbitrary bipartite state $\ket{X}_{A_1...A_mB_1...B_n}$,
if the von Neumann entropy of either of the subsystems is
$n (\le m)$, then it can be used to teleport faithfully an
arbitrary $n$-qubit state. Conversely, by applying
arbitrary unitary operators ${\cal U}^{-1}_A$ to the state
$\ket{\Lambda}_{AB}$ given in Eq. (\ref{Lambda}), one can
generate any number of states which support $n$-qubit
teleportation. Indeed, all of the special channels proposed
in the literature can be obtained this way
\cite{Yeo-06,Chen-06,Lee-02,Rigolin-05,Agrawal-06,Mura-08,Zha-07}.

Next we consider the non-maximally entangled case
 \beq
 E_{AB}<n.
 \eeq
In this case, perfect teleportation of an arbitrary
$n$-qubit state will succeed only probabilistically.
Nevertheless it may still be used to teleport a state of
$d(<n)$ qubits. Let
 \beq
 \ket{X'}_{AB}
 ={\cal U}_B\ket{X}_{AB},
 \eeq
where is ${\cal U}_B$ is an unitary operator in $H_B$ which
maximizes the value of $d (\le E_{AB})$ in the following
expression,
 \beq
 \rho'_B=\textrm{Tr}_{A}\ket{X'}_{AB}\bra{X'}_{AB}=
 \eta_{B'} \frac{1}{2^d}\prod_{i=1}^d I_{B_i},
 \label{condition}
 \eeq
where $I_{B_i}$ is the $2\times 2$ identity operator for
qubit $B_i$, and $\eta_{B'}$ is the density matrix of the
qubits in $B'=\{B_{d+1},...,B_n\}$. (Note that some
relabelling may be required.) Then using an unitary
operator ${\cal U}_A$, Alice can transform $\ket{X'}_{AB}$
into a state
 \beq
 \ket{\lambda}_{AB}=\Big(\prod_{i=1}^d
 \ket{\phi^k}_{A_iB_i}\Big)
 \ket{\varphi}_{A'B'},
 \eeq
where $k\in\{1,2,3,4\}$, $\ket{\phi^k}$ are Bell states,
$A'=\{A_{d+1},...,A_m\}$, and
 \beq
 \textrm{Tr}_{A'}\ket{\varphi}_{A'B'}\bra{\varphi}_{A'B'}
 =\eta_{B'}.
 \eeq
This is so because $\ket{X'}_{AB}$ and $\ket{\lambda}_{AB}$
have the same reduced density matrix $\rho'_B$ on Bob's
side \cite{Hughston-93}. Hence
 \beq
 \ket{\lambda}_{AB}={\cal U}_A{\cal U}_B\ket{X}_{AB},
 \label{UAUB}
 \eeq
and Alice can employ the $d$ shared Bell states in
$\ket{\lambda}_{AB}$ to teleport an arbitrary $d$-qubit
state to Bob. When $d=0$, not even a single qubit can be
teleported perfectly.

In the maximally entangled case $(E_{AB}=n)$, we have
${\cal U}_B=I$, and $d=n$ in Eq. (\ref{condition}). So the
general condition for faithful teleportation can be stated
as follows. If there exists an unitary operator ${\cal
U}_B$ such that Eq. (\ref{condition}) holds, then
$\ket{X}_{AB}$ can be used to teleport faithfully an
arbitrary state containing $d$ qubits. This condition is
also necessary. The following general protocol works for
any arbitrary channel state $\ket{X}_{AB}$ satisfying Eq.
(\ref{condition}): (1)Bob calculates ${\cal U}_B$ which
determines the maximum number ($d$) of qubits that can be
teleported, and Alice calculates ${\cal U}_A$. (2)Alice and
Bob apply ${\cal U}_A$ and ${\cal U}_B$ respectively to the
qubits in their control. (3)Then Alice can use the
resulting $d$ shared Bell states to teleport an arbitrary
$d$-qubit state to Bob.

Note that, if $E_{AB}=n$, then Alice is required to perform
one $m$-qubit operation and $n$ Bell state measurements,
and Bob is required to make $n$ single qubit operations at
most. In other protocols proposed for multiqubit
teleportation, Alice is required to perform operations
which are more complex than ours. For example, in
\cite{Rigolin-05,Agrawal-06,Yeo-06,Chen-06,Mura-08}, Alice
is required to perform a joint operation involving $(m+n)$
qubits. In our case this operation is broken down to a
series of operations involving smaller number of qubits,
which are much easier to perform experimentally. On the
other hand, if $E_{AB}<n$, then Bob may be required to
perform a $n$-qubit operation ${\cal U}_B$ as well. The
generalization to $m<n$ should be straightforward by now.

Using similar arguments, one can show that teleportation
could also be performed without making Bell state
measurements. For simplicity, we will show how it works for
teleporting a one-qubit state $\ket{\Psi}_1$. The original
procedure corresponds to Eq. (\ref{teleport-N}) with $N=1$.
Let us replace the four Bell states
$\{\ket{\phi^1},\ket{\phi^2},\ket{\phi^3},\ket{\phi^4}\}$
by the four orthogonal product states
$\{\ket{\chi^i}\}=\{\ket{11},\ket{10},\ket{01},\ket{00}\}$
respectively. The result is
 \beq
 \ket{\xi}_{1ab}
 =-{1\over 2}\Big(\sum_{i=1}^4\ket{\chi^i}_{1a}U^i_{b}\Big)
 \ket{\Psi}_{b}.
 \eeq
The reduced density matrix on Bob's side is
 \beq
 \textrm{Tr}_{1a}(\ket{\xi}_{1ab}\bra{\xi}_{1ab})=I_b/2,
 \eeq
which is the same as that of the joint initial state
$\ket{\Psi}_1\ket{\phi^1}_{ab}$. Therefore, again by the
freedom of purification \cite{Hughston-93},
$\ket{\xi}_{1ab}$ is related to
$\ket{\Psi}_1\ket{\phi^1}_{ab}$ by an unitary
transformation ${\cal U}_{1a}$ on Alice's side,
 \beq
 \ket{\xi}_{1ab}={\cal U}_{1a}
 \Big(\ket{\Psi}_1\ket{\phi^1}_{ab}\Big).
 \eeq
It can be shown that (apart from an unimportant phase),
 \beq
 {\cal U}_{1a} = H_a C^a_1,
 \eeq
where $H_a$ is the Hadamard operator, and $C^a_1$ is the
controlled-not operator with qubit 1 as the target. Hence
the standard teleportation procedure can be replaced by the
following alternative scheme: (1)Alice applies the unitary
operator ${\cal U}_{1a}$ on qubits 1 and a. (2)She then
individually measures the states of these qubits in the
basis $\{\ket{0},\ket{1}\}$, and sends the outcomes ($i$)
to Bob. (3)Bob recovers the original state by applying the
correctional unitary operator $U^i_b$ to his qubit. Thus
the Bell measurement in the original protocol
\cite{Bennett-93} can be replaced by the unitary operation
$U_{1a}$ plus two single-qubit measurements. It turns out
that this protocol is equivalent to the mysterious looking
quantum computing circuit devised by Brassard et al.
\cite{Brassard-98}. In a way, our derivation explains why
there must exist an unitary quantum circuit for
teleportation, even thought the original proposal of
Bennett et al. used Bell state measurement.

Finally, as a simple demonstration, let us take the channel
state to be the $N$-qubit GHZ state
 \beq
 \ket{\textrm{GHZ}}_{1...N}=\frac{1}{\sqrt{2}}
 (\ket{0...0}_{1...N}+\ket{1...1}_{1...N}).
 \eeq
No matter how the qubits are partitioned between Alice and
Bob (provided that each party gets at least one qubit), the
entropy of entanglement $E_{AB}=1$, so it can be used to
teleport a one-qubit state at most. If Alice holds qubits
$\{1,...,N-1\}$ and Bob holds the last qubit $N$, then the
entanglement between the two subsystems is maximal. It
follows from Eq. (\ref{UA}) that there exists an unitary
operator ${\cal U}_A$ such that
 \beq
 {\cal U}_A\ket{\textrm{GHZ}}_{1...N}=\ket{\phi^4}_{1N}
 \prod_{i=2}^{N-1}\ket{0}_i.
 \eeq
It is easy to show that ${\cal U}_A$ is just a series of
controlled-not operators $C^1_i$:
 \beq
 {\cal U}_A=\prod_{i=2}^{N-1}C^1_i.
 \eeq
In this special case, Alice alone can transform
$\ket{\textrm{GHZ}}_{1...N}$ into the desired form given in
Eq. (\ref{Lambda}). In general, if Alice holds qubits
$\{1,...,m\}$ and Bob holds qubits $\{m+1,...,N\}$, the two
subsystems are not maximally entangled. Then according to
Eq. (\ref{UAUB}), both ${\cal U}_A$ and ${\cal U}_B$ are
required, i.e.,
 \beq
 {\cal U}_A{\cal U}_B\ket{\textrm{GHZ}}_{1...N}
 =\ket{\phi^4}_{1N}
 \prod_{i=2}^{N-1}\ket{0}_i,
 \eeq
where
 \beqa
 {\cal U}_A&=&\prod_{i=2}^{m}C^1_i,\\
 {\cal U}_B&=&\prod_{i=m+1}^{N-1}C^N_{i}.
 \eeqa
In both cases, Alice and Bob share one Bell state, so the
given channel can be used to teleport a single-qubit state
only.

In summary, we have considered issues related to faithful
teleportation when the given channel is an arbitrary
$(m+n)$-qubit state $\ket{X}_{AB}$ partitioned between
Alice and Bob in the $(m,n)$ manner. A general condition,
Eq. (\ref{condition}), is presented which allows one to
judge if this channel can support faithful teleportation of
an arbitrary $d$-qubit state, where $d\le
\textrm{min}(m,n)$. We construct a general protocol which
is applicable for any channel states satisfying this
condition. Many special-case protocols have been proposed
in the literature
\cite{Rigolin-05,Agrawal-06,Yeo-06,Chen-06,Mura-08}.
Compared with which, ours is easier to implement
experimentally. For example, in the $d=n$ case, the most
complex operation in our protocol involves $m$ qubits,
while the others involve $(m+n)$ qubits.





\begin{thebibliography}{}

\bibitem{Bennett-93} C. H. Bennett, G. Brassard, C. Cr\'epeau,
R. Josza, A. Peres, and W. K. Wootters, Phys. Rev. Lett.
\jo{70}{1895}{1993}.

\bibitem{Marcikic-03} I. Marcikic, H. de Riedmatten, W.
Tittel, H. Zbinden, and N. Gisin, Nature (London)
\jo{421}{509}{2003}.

\bibitem{Riebe-04} M. Riebe, H. Hffner, C. F. Roos, W.
Hnsel, J. Benhelm, G. P. T. Lancaster, T. W. Krber, C.
Becher, F. S. Kaler, D. F. V. James, and R. Blatt, Nature
(London) \jo{429}{734}{2004}.

\bibitem{Barrett-04} M. D. Barrett, J. Chiaverini, T.
Schaetz, J. Britton, W. M. Itano, J. D. Jost, E. Knill, C.
Langer, D. Leibfried, R. Ozeri, and D. J. Wineland, Nature
(London) \jo{429}{737}{2004}.

\bibitem{Zhao-04} Z. Zhao, Y. A. Chen, A. N. Zhang, T.
Yang, H. J. Briegel, and J. W. Pan, Nature (London)
\jo{430}{54}{2004}.

\bibitem{Ursin-04} R. Ursin, T. Jennewein, M. Aspelmeyer,
R. Kaltenbaek, M. Lindenthal, P. Walther, and Z. Zeilinger,
Nature (London) \jo{430}{849}{2004}.

\bibitem{Chuang-02} See, e.g., M. A. Nielsen and I. L. Chuang,
\textit{Quantum Computation and Quantum Information}
(Cambridge University Press 2000).

\bibitem{Yeo-06} Y. Yeo and W. K. Chua, Phys. Rev. Lett.
\jo{96}{060502}{2006}.

\bibitem{Chen-06} P. X. Chen, S. Y. Zhao, and G. C. Guo,
Phys. Rev. A \jo{74}{032324}{2006}.

\bibitem{Lee-02} J. Lee, H. Min, and S. D. Oh, Phys. Rev. A
\jo{66}{052318}{2002}.

\bibitem{Rigolin-05} G. Rigolin, Phys. Rev. A \jo{71}{032303}{2005}.

\bibitem{Agrawal-06} P. Agrawal and A. Pati, Phys. Rev. A
\jo{74}{062320}{2006}.

\bibitem{Mura-08} S. Muralidharan and P. K. Panigrahi,
Phys. Rev. A \jo{77}{032321}{2008}.

\bibitem{Zha-07} X. W. Zha and H. Y. Song, Phys. Lett. A
\jo{369}{377}{2007}.

\bibitem{Bennett-96} C. H. Bennett, H. J. Bernstein, S.
Popescu, and B. Schumacher, Phys. Rev. A
\jo{53}{2046}{1996}.

\bibitem{Popescu-97} S. Popescu and D. Rohrlich,
Phys. Rev. A \jo{56}{R3319}{1997}.

\bibitem{Hughston-93} L. P. Hughston, r. Jozsa, and W. K.
Wootters, Phys. Rev. A \jo{183}{14}{1993}.



\bibitem{Brassard-98} G. Brassard, S. Braunstein, and R. Cleve,
Physica D \jo{120}{43}{1998}; e-print
arXiv:quant-ph/9605035.


\end{thebibliography}
\end{document}